\documentclass{IOS-Book-Article}

\usepackage{mathptmx}
\usepackage{graphicx}
\usepackage{xcolor}
\usepackage{color}
\usepackage{listings}
\usepackage{bm}
\usepackage{url}
\usepackage{booktabs}
\usepackage{xspace}
\usepackage{subcaption}
\usepackage[binary-units=true,per-mode=symbol]{siunitx}

\def\hb{\hbox to 10.7 cm{}}

\newcommand{\propagate}{{\em propagate}}
\newcommand{\collide}{{\em collide}}
\newcommand{\AoS}{{\em AoS}}
\newcommand{\SoA}{{\em SoA}}
\newcommand{\CSoA}{{\em CSoA}}
\newcommand{\CAoSoA}{{\em CAoSoA}}

\newcommand{\ie}{i.e.,\xspace}

\begin{document}

\pagestyle{headings}
\def\thepage{}

\begin{frontmatter}              % The preamble begins here.

%\pretitle{Pretitle}
\title{Energy-efficiency evaluation of Intel KNL for HPC workloads
%cache and flat modes for HPC workloads
}

%\markboth{}{September 2016\hb}
%\subtitle{Subtitle}

\author[A]{\fnms{Enrico} \snm{Calore}%
\thanks{Corresponding Author: Enrico Calore, University of Ferrara and INFN,
Via Saragat 1, I-44122 Ferrara Italy; E-mail: enrico.calore@unife.it.}},
\author[A]{\fnms{Alessandro} \snm{Gabbana}}
\author[A]{\fnms{Sebastiano Fabio} \snm{Schifano}}
and
\author[A]{\fnms{Raffaele} \snm{Tripiccione}}

\runningauthor{E. Calore et al.}
\address[A]{University of Ferrara and INFN, Italy}

\begin{abstract}
Energy consumption is increasingly becoming a limiting factor to the design of 
faster large-scale parallel systems, and development of energy-efficient and 
energy-aware applications is today a relevant issue for HPC code-developer 
communities.
In this work we focus on energy performance of the {\em Knights Landing} (KNL)
Xeon Phi, the latest many-core architecture processor introduced by Intel for  
the HPC market.
We take into account the 64-core Xeon Phi 7230, and analyze its energy 
efficiency using both the on-chip MCDRAM and the regular DDR4 system memory as
main storage for the application data domain.
As a benchmark application we use a Lattice Boltzmann code heavily optimized for 
this architecture, and implemented using different memory data layouts to store 
the data-domain.
We then assess the energy consumption using different data-layouts,  
memory configurations (DDR4 or MCDRAM), and number of threads per core.
\end{abstract}

\end{frontmatter}
%\markboth{July 2017\hb}{July 2017\hb}
\thispagestyle{empty}

\section{Introduction}
\label{sec:intro}

Energy consumption is quickly becoming one of the critical issues in modern HPC
systems, and HPC facilities are constantly trying to increase their 
energy-efficiency, in order to hold down the total cost of ownership, 
increasingly given by the electricity bill.
Processors and accelerators are the main sources of power drain in computing 
systems~\cite{powerpack} and thus assessing their energy-efficiency is of 
paramount importance for the design of efficient parallel systems.
Hardware manufacturers are clearly trying to improve the energy-efficiency 
increasing the peak FLOP/watt ratio of their architectures~\cite{hpc-trends}.
Beside of this, it is also relevant to analyze and profile the hardware 
energy-performance running real-applications, since high 
energy-efficient processors may behave highly inefficiently when running codes 
unable to exploit a large fraction of their peak performance.
For this reason in this work we study the energy-efficiency of the Intel Knights
Landing (KNL) architecture, using as a benchmarking code a real HPC application, 
heavily optimized for several architectures and used for production runs of 
fluid-dynamics simulations based on Lattice Boltzmann methods.
This application is a good representative of a wider class of lattice based 
stencil-codes, including also HPC \textit{Grand Challenge} applications such as
Lattice Quantum Chromodynamics (LQCD)~\cite{bernard,bilardi,se4hpcs15,lqcdoacc17}. 

The Intel KNL features a 16GB on-package HBM (High Bandwidth Memory) namely 
MCDRAM (Multi-Channel DRAM) which resides on the CPU chip, close to the 
processing cores.
This memory can be configured as a wide cache, interposed between the cores and 
the DDR4 system memory, or as a separate addressable space.

In this work we evaluate the impact on the energy consumption required by the
KNL to complete different kernels of our application, using different number of
threads per core, different memory configurations (i.e. MCDRAM and DDR4), and 
different memory data layouts.
%
% Questo vediamo se metterlo o meno... magari qualche cosa si puo' anche
% mettere, tenendoci qualche risultato da parte...
%
We then compare the results with recent NVIDIA GPUs. 
%
% possiamo qui mettere i risultati di Haswell e K80 e nella versione journal il Broadwell e la P100 ??
%
To benchmark different processors we adopt different code implementations of
the same LBM application, each apecifically optimized for a different target 
system.

\section{Lattice Boltzmann Application}
\label{sec:lbm}

Behaviour of flows are widely described using the Lattice Boltzmann 
method (LB)~\cite{sauro} based on the synthetic dynamics of {\em populations} 
sitting at the sites of a discrete lattice. 
These methods, discrete in position and momentum spaces, {\em propagate} populations 
from lattice-sites to lattice-sites, and then {\em collide} them changing their 
values accordingly. 
In this work we consider a state-of-the-art $D2Q37$ LB model   
with 37 populations per lattice-site, that correctly reproduces
the thermo-hydrodynamical evolution of a fluid in two dimensions, and
enforces the equation of state of a perfect gas ($p = \rho T$)~\cite{JFM,POF}. 
This model has been extensively used for large scale simulations of convective 
turbulence~\cite{noi1} on HPC systems~\cite{iccs10}.

A Lattice Boltzmann simulation starts with an initial assignment of the 
populations values, and then for each lattice-point applies in sequence 
two kernel functions as many time-steps as needed. 
The first kernel, called {\sf propagate}, moves populations across lattice sites 
collecting at each site all populations that will interact at the next step 
({\sf collide}) according to an appropriate stencil depending on the LB model used.
This kernel performs only a large number of sparse memory accesses, and for this 
reason is strongly memory-bound.
The latter, called {\sf collide}, uses as input the populations gathered by the 
previous {\sf propagate} kernel, and performs all the mathematical steps associated 
to the computation of the new population values. This function is strongly 
compute-bound making heavy use of the floating-point units of the processor.
These two kernels take a large fraction of the simulation execution time.

In the last years several implementations of this model have been developed, 
and used both for convective turbulence studies~\cite{noi2}, and as benchmarks 
to profile performance of recent developed HPC hardware architectures based on 
commodity CPUs and GPUs~\cite{ppam11,europar14,uchpc15,ccpe16,parco16}.
In this work we use an implementation initially developed for Intel CPUs~\cite{caf13}, 
and later ported and optimized for the Intel {\em Knights Corner} (KNC) processor~\cite{iccs13,ppam15}.

\section{The Knights Landing Architecture}
\label{sec:knl-architecture}

The {\em Knights Landing} (KNL) is the first generation of self-bootable 
processor based on the {Many Integrated Cores} (MIC) architecture developed 
by Intel.
The processor integrates an array of 64, 68 or 72 cores together with four 
high speed {\em Multi-Channel DRAM} (MCDRAM) memory banks, providing an aggregate 
raw peak bandwidth of more than \SI{450}{\giga\byte\per\second}~\cite{stream}. 
It also integrates 6 DDR4 channels supporting up to \SI{384}{\giga\byte} of memory 
with a peak raw bandwidth of \SI{115.2}{\giga\byte\per\second}.
Two cores are bonded together into a tile sharing an L2-cache of \SI{1}{\mega\byte}. 
Tiles are connected by a 2D-mesh of rings and can be clustered in several NUMA 
configurations.
In this work we only run using the {\em Quadrant} cluster configuration where  
the array of tiles is partitioned in four quadrants, each connected to one 
MCDRAM controller.
This configurations is the one recommended by Intel to run applications using the KNL 
as a symmetric multi-processor, as it reduces the latency of L2-cache misses, and the 
4 blocks of MCDRAM appear as contiguous block of memory addresses~\cite{colfax-knl-numa}. 

MCDRAM on a KNL can be configured at boot time as \textit{Flat}, \textit{Cache} 
or \textit{Hybrid} mode.
The \textit{Flat} mode configures the MCDRAM as a separate addressable memory, 
\textit{Cache} mode configures the MCDRAM as a last-level cache. The \textit{Hybrid} mode, 
allows to use the MCDRAM partly as addressable memory and partly as last-level 
cache~\cite{colfax-knl-mcdram}.
In this work we only consider \textit{Flat} and \textit{Cache} configurations.

KNL allows to exploit two levels of parallelism: the first is the {\em task parallelism} 
built onto the array of cores, and the latter is the {\em data parallelism} 
using the AVX 512-bit vector (SIMD) instructions.
Each core has two out-of-order vector processing units (VPUs) and supports the 
execution of up to 4 threads.
The KNL has a peak theoretical performance of 6 TFlops in single precision 
and 3 TFlops in double precision, and the typical thermal design power (TDP) 
is \SI{215}{\watt}, including MCDRAM memories.
For more details on KNL architecture see~\cite{sodani16}.

\section{Measuring Energy Consumption on the KNL}
\label{sec:knl-energy}

As other Intel processors -- starting from Sandy Bridge architecture onward --
the KNL integrates power meters and a set of {\em Machine Specific 
Registers} (MSR) that can be read through the {\textit{Running Average 
Power Limit} (RAPL) interface.
%
%This interface allows to access a set of counters providing energy consumption 
%information.
In this work we use the \textit{PACKAGE\_ENERGY} and \textit{DRAM\_ENERGY} 
counters to monitor respectively the energy consumption of the processor Package
(accounting for: Cores, Caches and MCDRAM) and the DRAM memory system.

A popular way to access these counters is through a library named 
PAPI~\cite{papi} providing a common API for energy/power readings for 
different processors, partially hiding architectural details.
The use of this technique to read energy consumption figures from Intel 
processors is a consolidate practice~\cite{papi-energy}, validated by several 
third parties studies~\cite{power-haswell,papi-dram-validation}.

On top of the PAPI library we have developed a custom library to manage power/energy 
data acquisition from hardware registers. 
It allows benchmarking codes to directly start and stop measurements, using 
architecture specific interfaces, such as RAPL for Intel CPUs and the 
\textit{NVIDIA Management Library} (NVML) for NVIDIA GPUs~\cite{ccpe17}.
Our library also lets benchmarking codes to place markers in the data stream in 
order to have an accurate time correlation between the running kernels and the 
acquired power/energy values. 
Our code, exploiting the PAPI library, is available for download as Free
Software~\cite{papi-reader}.

We use the \textit{energy-to-solution} ($E_s$) as metric to measure the 
energy-efficiency of processors running our application. 
This is defined as product of \textit{time-to-solution} ($T_s$) and average 
power ($P_{avg}$) drained while computing the workload: 
$E_s = T_s \times P_{avg}$.
To measure $P_{avg}$ we read the RAPL hardware counters thanks to our wrapper
library, and then convert readout values to Watt.
The Package and DRAM power drains can then be summed to obtain the
overall value or analyzed separately.

In the following, we measure this metric for the two most time consuming kernels
of our application, {\propagate} which is strongly memory-bound, and {\collide} 
which is strongly compute-bound.

\section{Experimental Results}
\label{sec:knl-measure}

The LB application described in Sec.~\ref{sec:lbm} has been originally 
implemented using the {\AoS } (\textit{Array of Structure}) data layout, showing
a satisfactory performance on CPU processors.
While porting the application to GPU devices, a data layout re-factoring has been 
implemented to fully exploit the parallelism of GPUs, leading to an
implementation based on the {\SoA } (\textit{Structure of Array}) data 
layout.

More recently, two slightly more complex data layouts have been 
introduced~\cite{ijhpca17} in the quest of a single data structure to be used 
for a portable implementation, allowing good performances on most architectures.
These data structures, named: {\CSoA } (\textit{Clustered Structure of Array}) 
and the {\CAoSoA } (\textit{Clustered Array of Structure of Array}), allow to 
exploit vectorization on both CPUs and GPUs, still guaranteeing efficient data 
memory accesses and vector processing, for both of our application's critical 
kernels.

In this work we test all of the above data layouts on the KNL
architecture, measuring the energy consumption of both the Package and DRAM.
Moreover we run with various number of threads and with different memory 
configurations: i) allocating the lattice only in the DRAM using the 
Flat/Quadrant configuration; ii) allocating the lattice only in the MCDRAM 
using the Flat/Quadrant configuration; or iii) allocating the lattice only in 
the DRAM, but using the MCDRAM as a last level cache, using the Cache/Quadrant 
configuration.
In the latter case we use a bigger lattice, which could not fit in the cache, 
sand for this reason we report all the results normalized per lattice site.
Our aim is to analyze and highlight possible differences in performance, average 
power, and energy-efficiency.

In Fig.~\ref{fig:knl-propagate} and Fig.~\ref{fig:knl-collide} we show the
results of our tests, where Power and Energy values account for the sum of 
Package and DRAM contributions and Energy is normalized per lattice site.

\begin{figure}
  \begin{subfigure}[b]{\textwidth}
    \includegraphics[width=\textwidth]{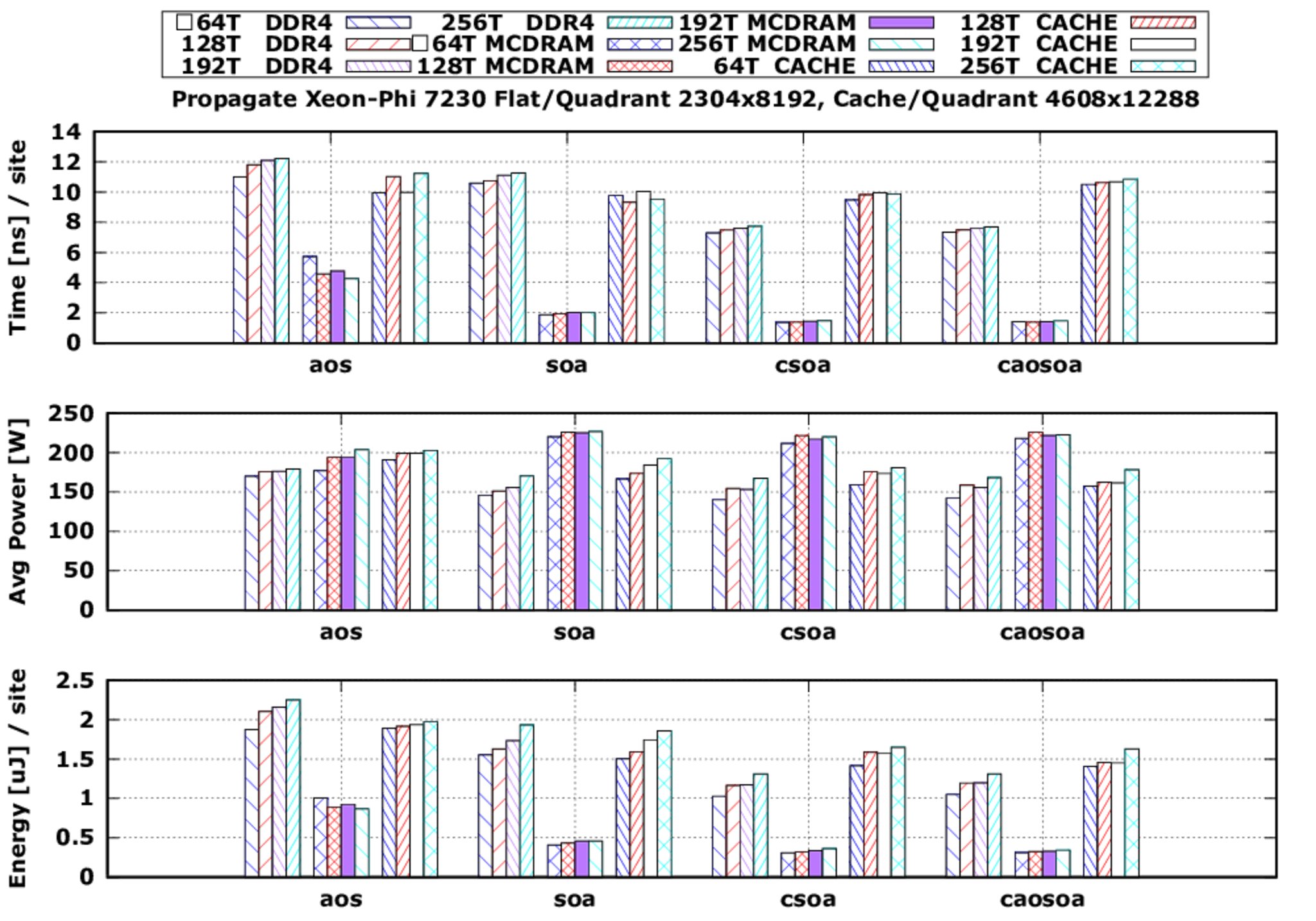}
    \caption{{\propagate} function: 
    we report $T_s$ in nano-seconds per site (Top), $P_{avg}$ in Watt (Middle) 
    and $E_s$ in micro-joules per site (Bottom).}
    \label{fig:knl-propagate}
  \end{subfigure}
% \end{figure}

% \begin{figure}
\begin{subfigure}[b]{\textwidth}
    \includegraphics[width=\textwidth]{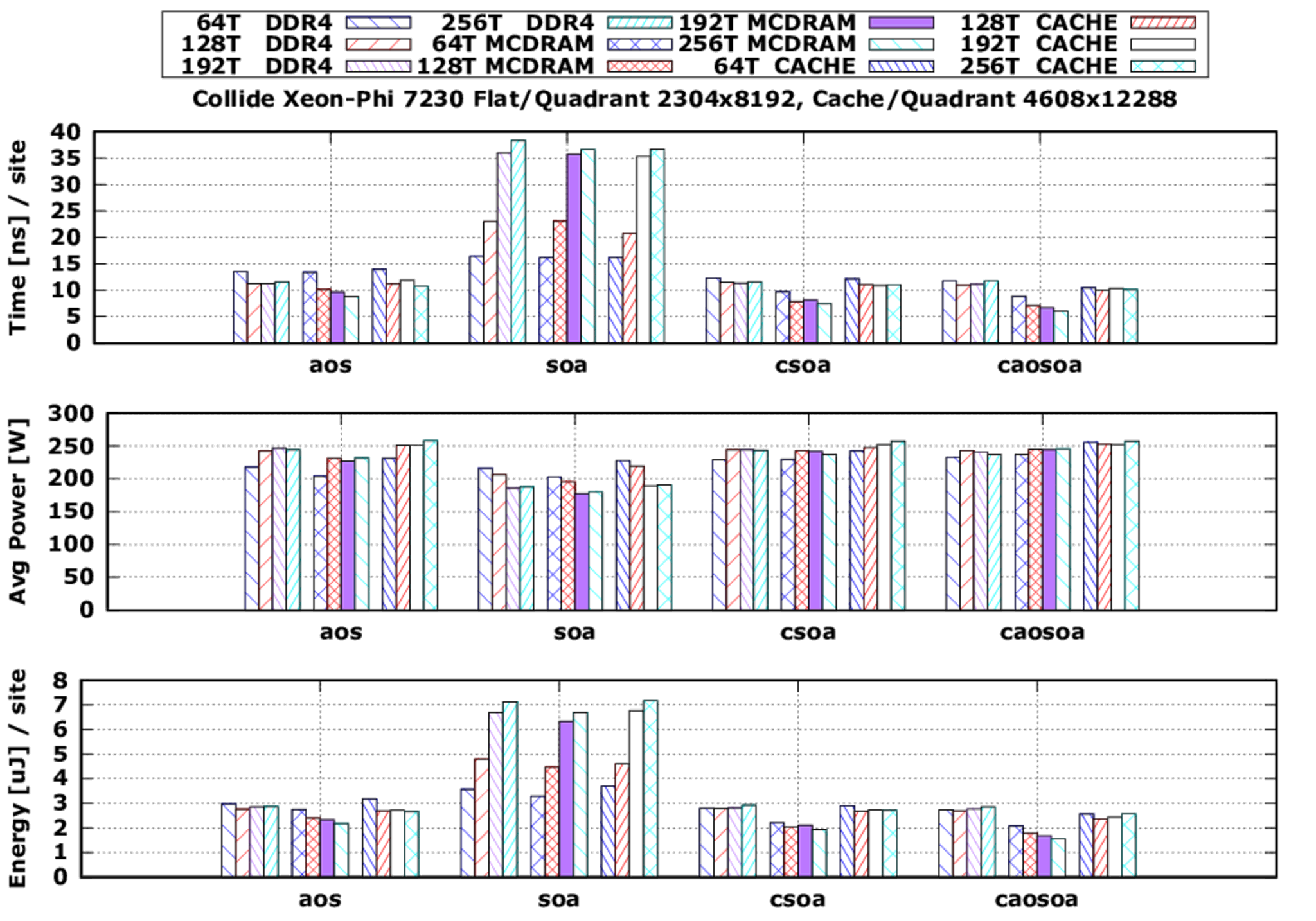}
    \caption{{\collide} function:  
    we report $T_s$ in nano-seconds per site (Top), $P_{avg}$ in Watt (Middle) 
    and $E_s$ in micro-joules per site (Bottom).}
    \label{fig:knl-collide}
  \end{subfigure}
  
\caption{All the performed tests for the three memory configuration (DDR4 Flat, 
MCDRAM Flat and Cache) and for different thread numbers (64, 128, 192 and 256).
We report three metrics: the time-to-solution ($T_s$), the average Power drain 
($P_{avg}$) and the energy-to-solution ($E_s$).
In Fig.~\ref{fig:knl-propagate} for the {\propagate} function and in 
Fig.~\ref{fig:knl-collide} for the {\collide}.}
  
\end{figure}

% PROP
%%%%%%%%%%%%%%%%%%%%%%%%%%%%%%%%%%%%%%%%%%%%%%%%%%%%%%%%%%%%%%%%%%%%%%%%%%%%%%%

Concerning the {\propagate} function, as shown in Fig.~\ref{fig:knl-propagate},
using Flat-mode and allocating the lattice in the MCDRAM memory, the maximum 
bandwidth using the {\AoS} data layout is \SI{138}{\giga\byte\per\second}, 
while using {\SoA} it can be increased to \SI{314}{\giga\byte\per\second} (\ie 
$2.3 \times$), and then can be further increased using {\CSoA} layout, reaching a
peak bandwidth of \SI{433}{\giga\byte\per\second} ($3.1\times$ wrt 
{\AoS}).
When using just the regular DRAM, bandwidth drops for all the data layouts:
\SI{51}{\giga\byte\per\second} for {\AoS}, \SI{56}{\giga\byte\per\second} for
{\SoA}, and \SI{81}{\giga\byte\per\second} for {\CSoA}.
When using the Cache-mode with a bigger lattice size, which do not fits in the
MCDRAM, we measure a quite constant bandwidth of $59$, 
$60$ and \SI{62}{\giga\byte\per\second} respectively for {\AoS}, {\SoA} and
{\CSoA}.
The {\CAoSoA } data layout on the other side, do not provides any benefit over
the {\CSoA} for the {\propagate} function, but as we see in 
Fig.~\ref{fig:knl-collide}, it is beneficial for the performance of the 
{\collide}.

% With reference to the lower plots of Fig.~\ref{fig:knl-propagate}, the average 
% Power drain and corresponding {\em Energy to Solution} $E_S$ are given as the 
% sum of Package and DRAM contributions, and $E_S$ is normalized per the lattice 
% size.
%
We record the best $E_S$ using Flat-MCDRAM configuration and the {\CSoA} 
data layout with a result of $\approx 2.5\times$ energy saving wrt using 
the {\AoS} data layout.

From the point of view of all the evaluated metrics, using just 64 threads is 
the best choice, since it gives the best performance plus the lowest Power drain 
and Energy consumption.
This is justified by the fact that the {\propagate} function is completely 
memory-bound and 64 threads are enough to keep fully busy the memory controllers.

% COLL
%%%%%%%%%%%%%%%%%%%%%%%%%%%%%%%%%%%%%%%%%%%%%%%%%%%%%%%%%%%%%%%%%%%%%%%%%%%%%%%
  
Concerning the {\collide} function, similar plots are reported in 
Fig.~\ref{fig:knl-collide}, where again the average Power drain and $E_S$ are 
given as the sum of Package and DRAM contributions and $E_S$ is normalized per  
lattice size.
For Flat-MCDRAM configuration, performance increases when changing the data 
layout from {\AoS} to {\CSoA} and in this case it can be further increased  
changing to {\CAoSoA }.
On the other side, the worst performance is shown using {\SoA}, since 
vectorization can not be exploited successfully in this case, and moreover 
memory-alignment can not be granted.
When using {\CAoSoA } we measure a sustained performance of $\approx 1$Tflops, 
corresponding to $\approx 37\%$ of the raw peak performance of the KNL.
%and is in line with the maximum performance efficiency obtainable on other 
%processors and accelerators.
%
From our measures, the {\CAoSoA } layout gives the best $E_S$, 
$\approx 2\times$ lower wrt to the {\AoS}, both for performance and $E_S$.

%When using Cache configuration, performance becomes limited by memory bandwidth.

In contrast to the behaviour of the {\propagate}, for the {\collide} function 
$E_S$ decreases using more threads per CPU since this kernel is compute-bound.

%%%%%%%%%%%%%%%%%%%%%%%%%%%%%%%%%%%%%%%%%%%%%%%%%%%%%%%%%%%%%%%%%%%%%%%%%%%%%%%

For both functions we see that the $E_S$ differences, between the 
various tests performed, are mainly driven by $T_S$.
Despite the fact that average Power drain shows differences up to $\approx 30\%$,
it seems always convenient to choose the best performing configuration from the 
$T_S$ point of view, to obtain also the best energy-efficiency.

%%%%%%%%%%%%%%%%%%%%%%%%%%%%%%%%%%%%%%%%%%%%%%%%%%%%%%%%%%%%%%%%%%%%%%%%%%%%%%%

In Fig.~\ref{fig:knl-energy} we highlight the $E_S$ metrics for all the 
different data layouts, using the Flat configuration and thus using either the 
off-chip (DDR4) or the on-chip (MCDRAM) memory, to allocate the whole lattice.
Here we display separately the Package and DRAM contributions, where 
for each bar in the plot, Package energy is in the bottom and DRAM energy on top.
When using the MCDRAM, its energy consumption is accounted with the rest of the 
Package and thus what is displayed as DRAM energy is just due to the idle Power 
drain.

The main advantage in using the MCDRAM is clearly for the {\propagate} function
where we can save $\approx 2/3$ of the energy wrt the best performing test run 
using the DDR4 system memory (be aware of the different scales in the
y-axes of Fig.~\ref{fig:knl-propagate-energy}).
Anyhow, also for the {\collide} function, shown in Fig.~\ref{fig:knl-collide-energy}
we can halve the energy consumption.
In both the cases the energy saving is mainly given by the reduced execution 
time, since the DDR4 average Power drain accounts at most for $\approx 10\%$.

\begin{figure}
  \begin{subfigure}[b]{\textwidth}
 \includegraphics[width=\textwidth]{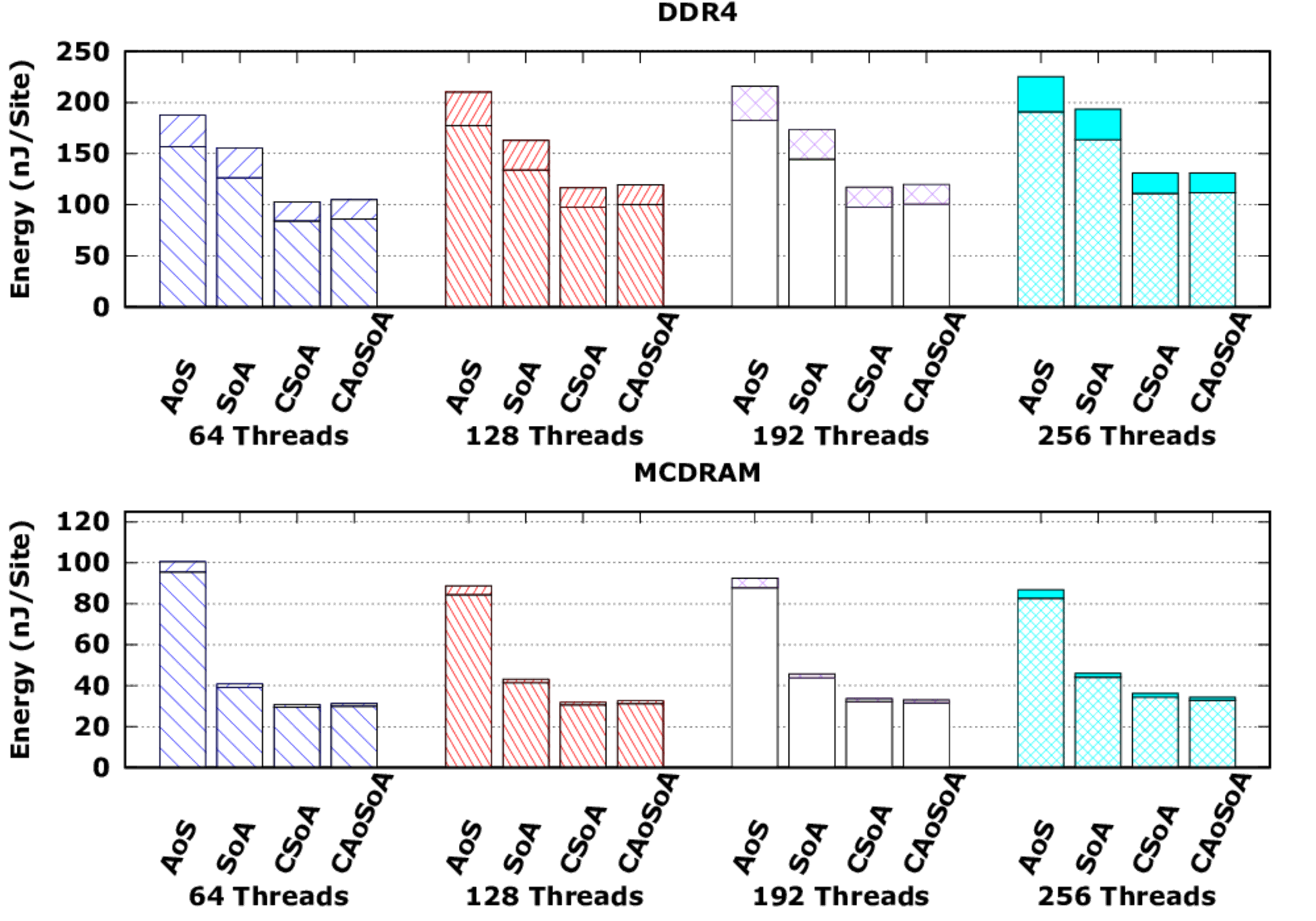}
 \caption{{\propagate} function. Flat configuration, using the DDR4 system 
 memory (Top) and the MCDRAM (bottom). We highlight to the reader the different
 scales on the y-axes.}
 \label{fig:knl-propagate-energy}
 \end{subfigure}
%\end{figure}

%\begin{figure}
\begin{subfigure}[b]{\textwidth}
  \includegraphics[width=\textwidth]{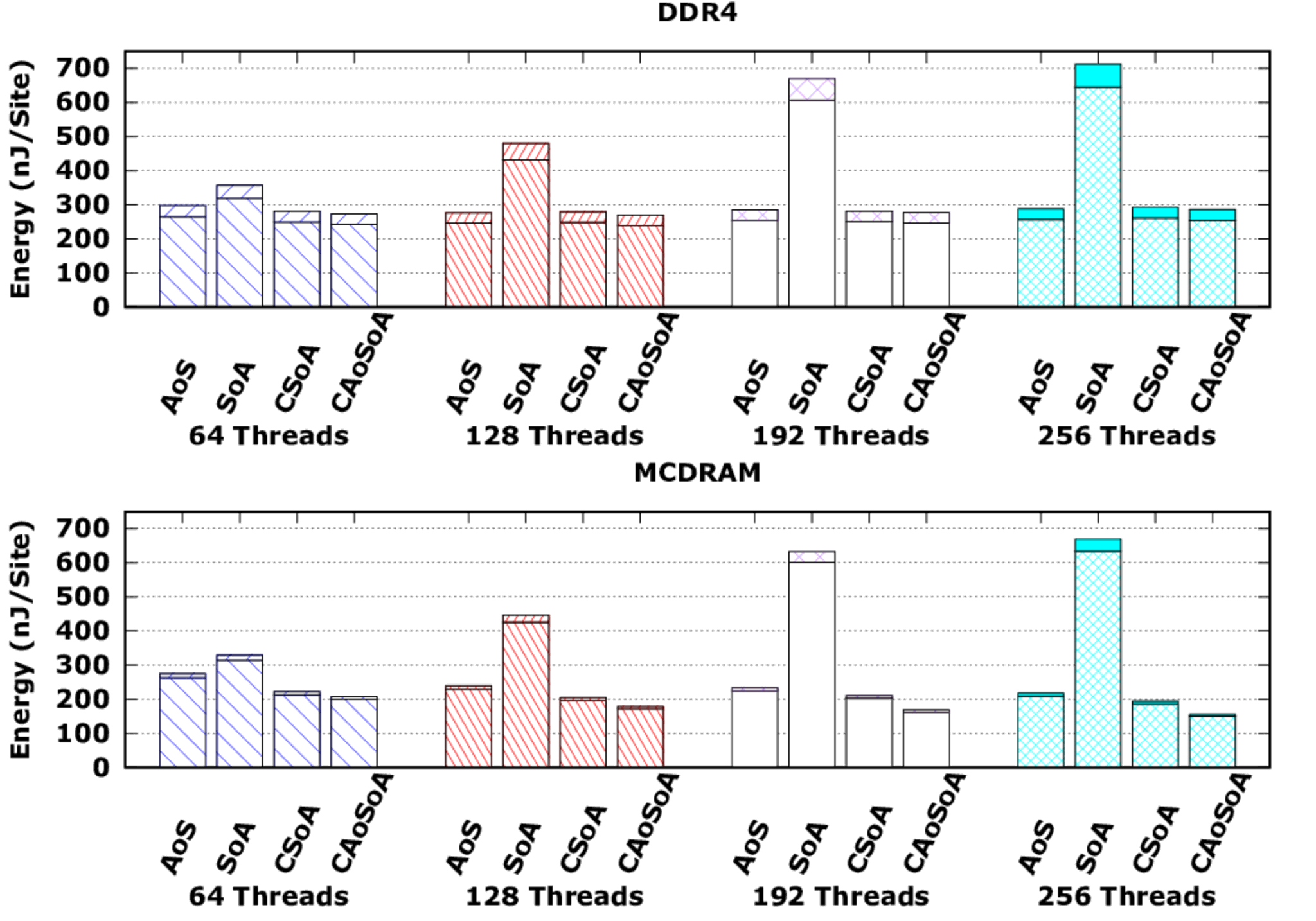}
  \caption{{\collide} function. Flat configuration, using the DDR4 system 
 memory (Top) and the MCDRAM (bottom).}
  \label{fig:knl-collide-energy}
 \end{subfigure}
 
 \caption{Energy consumption in nano-joules per lattice site, using different 
 memory data layouts and different number of threads.
 Each bar represent the Package Energy (bottom), plus the DRAM energy (top), 
 which is just the DRAM idle energy consumption, when using the MCDRAM).
 }
 \label{fig:knl-energy}
\end{figure}

\section{Comparison with other architectures}
\label{sec:comp}

In this section we comment on the energy-efficiency of the KNL, comparing it to 
other processors and accelerators.
We have run different implementations of the same code presented in Sec.~\ref{sec:lbm},
each optimized for a specific processor or accelerator, and we have measured the 
corresponding energy-to-solutions.
To read power and energy hardware counters we have used our custom wrapper code based 
on the PAPI library presented in Sec.~\ref{sec:knl-energy}.

  \begin{figure}[ht]
  \includegraphics[width=0.8\textwidth]{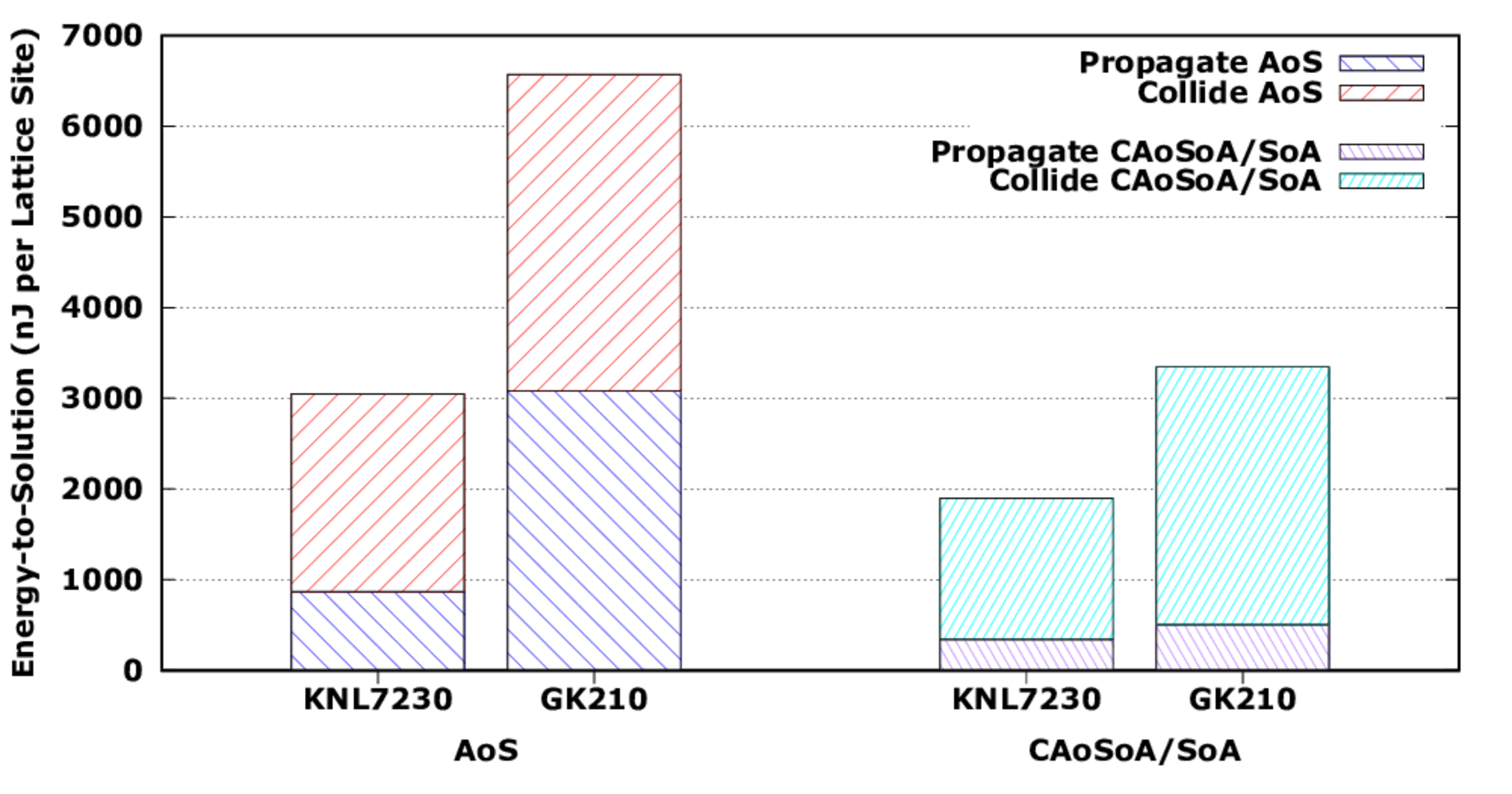}
  \caption{Energy-to-solution of the {\propagate} and {\collide} functions, using
  the {\AoS} data layout and run on the KNL and on an NVIDIA GK210 GPU, on the 
  left.
  The same while running with the most energy-efficient data layout (\ie 
  {\CAoSoA } for the KNL and {\SoA} for the GK210) on the right.}
  \label{fig:knl-comparison}
 \end{figure}
    
In Fig.~\ref{fig:knl-comparison} we compare the energy-efficiency of a KNL with 
that of a GK210 GPU hosted into an NVIDIA K80 board. We have configured the KNL 
in Flat mode, and used a lattice size that fits in the 16GB of the KNL's MCDRAM 
as well as in the on-board GPU memory.
For both the architectures we report the results using a naive {\AoS} 
implementation and using the respective best performing memory layout.

To use the {\AoS } data layout is sub-optimal for both the architectures, and 
the KNL exhibits a lower $E_S$ than the GPU by a factor $\approx 3x$ for both 
{\propagate} and {\collide} kernels.
Using the most energy-efficient data layout for both architectures -- \ie 
{\CAoSoA } for the KNL and {\SoA} for the GK210 -- the {\propagate} and the 
{\collide} kernels are respectively $\approx 31\%$ and $\approx 55\%$ more 
energy-efficient on the KNL compared to the GPU.
Using the Cache configuration for the KNL (not shown Fig.~\ref{fig:knl-comparison}), 
if the lattice fits into the cache, computing performance and energy-efficiency 
are the same measured for Flat mode.
For lattices which do not fit into the cache, both computing performance and 
energy-efficiency of the KNL drops to the level of standard multi-core Intel 
Xeon CPUs.

Considering the fact that the KNL architecture is more recent than the GK210 
used in this comparison, we can conclude that the energy-efficiency of Intel
Xeon Phi and of NVIDIA GPUs is comparable if data can fit in the KNL's MCDRAM.

\section{Conclusion and future works}
\label{sec:conclusion}
 
%% RICORDARSI DI CITARE:
%
% ppam17
%
% Exploring the Performance Benefit of Hybrid Memory System on HPC Environments
%
% Applying the Roofline Performance Model to the Intel Xeon Phi Knights Landing Processor
%

In this work we have investigated the energy-efficiency of the Intel KNL for 
Lattice Boltzmann applications, assessing the {\em energy to solutions} 
for the most compute relevant kernels, \ie {\propagate} and {\collide}.
Based on our experience in using the KNL, related to our application, and 
the experimental measures we have observed, some concluding remarks are in order:

i) applications previously developed for ordinary X86 multi-core CPUs can be 
easily ported and run on KNL processors;

ii) performance is strongly affected by the level of vectorization and core 
parallelism that applications are able to exploit, otherwise a drop to the level
of ordinary multi-core CPUs or even worst can be easily observed;

iii) for LB applications -- and for many others -- appropriate data layouts 
plays a relevant role to allow for vectorization and an efficient use of the 
memory sub-system, improving computing and energy efficiency;
%while average power does not change significantly;

iv) if application data fits within the MCDRAM, performance of KNL are very 
competitive with that of recent GPU accelerators in terms of both computing and 
energy-efficiency; unfortunately, if this is not the case, performance drops to 
levels of ordinary multi-core CPUs or even worst;
%since tools and operations (editing, 
%compilations, IO, etc.) not exploiting core and data parallelism run much slower 
%making difficult to use the KNL as standalone processor.

In the future, we plan to further investigate energy-efficiency of the KNL 
providing a more comprehensive comparison, taking into account more recent GPUs 
architectures (such as Pascal and Volta), and also exploring the possibility
to use DVFS to tune the KNL clock frequency.

{\scriptsize
\subsubsection*{Acknowledgements}
This work was done in the framework of the COKA, COSA projects of INFN, 
and the PRIN2015 project of MIUR. We would like to thank CINECA (Italy) for 
access to their HPC systems. AG has been supported by the EU Horizon 2020 
research and innovation programme under the Marie Sklodowska-Curie grant 
agreement No 642069.
}

\end{document}